\documentstyle[12pt,epsfig,amssymb]{article}
\begin{document}
\setlength{\baselineskip}{0.75cm}
\setlength{\parskip}{0.45cm}
\renewcommand{\theequation}{\arabic{section}.\arabic{equation}}
\begin{titlepage}
\begin{flushright}
\large
DO-TH 96/08\\ May 1996
\end{flushright}
\normalsize
\vspace{1.3cm}
\begin{center}
\LARGE
\hbox to\textwidth{\hss
{\bf Probing the Parton Densities of Virtual Photons} \hss}

\vspace{0.1cm}
\hbox to\textwidth{\hss
{\bf at ep Colliders}  \hss}

\vspace{1.7cm}
\large
M.\ Gl\"uck, E.\ Reya and M.\ Stratmann\\
\vspace{1cm}
Institut f\"ur Physik, Universit\"at Dortmund \\
D--44221 Dortmund, Germany
\end{center}
\vspace{2.0cm}
\underline{{\large{Abstract}}} \\ \\
\normalsize
\setlength{\baselineskip}{0.75cm}
\setlength{\parskip}{0.45cm}
The feasibility of an experimental determination of the parton
distributions of virtual photons at high energy $ep$ colliders is
studied in the context of the DESY-HERA collider. Recently proposed
parton densities of virtual photons are utilized to evaluate the
appropriate production processes and their relevant kinematical regions.
It is demonstrated that high $E_T$ jet production with $E_T\,\simeq 5-7\,
{\rm{GeV}}$ and $b\bar{b}$ production can probe the parton distributions
of virtual photons up to virtualities of a few ${\rm{GeV}}^2$. A useful
leading order parametrization of our proposed photonic parton densities,
suitable for further studies of these issues, is presented.
\end{titlepage}
\newpage
%
%
\section{Introduction}
The parton distributions of photons provide us with a unique opportunity
to study the underlying mechanisms responsible for the specific shapes
of hadronic parton distributions. This is due to the fact that in
contrast to the situation with nucleons or pions the photon provides us
with the possibility of investigating its parton distributions in a
{\underline{continuous}} range of masses (virtualities) $P^2=-p^2_{\gamma}$
with $p_{\gamma}$ denoting the four momentum of the photon emitted from,
say, an electron in an $e^+e^-$ or $ep$ collider. (In the latter case it
is common to use $Q^2=-q^2$ instead of $P^2$ for denoting the photon's
virtuality, but we prefer $P^2$ according to the original notation used
in $e^+e^-$ annihilations where it refers to the virtuality of the probed
virtual target photon \cite{ref1}.) This is particularly interesting for
models based on some specific notion concerning the nonperturbative
origin of the hadronic parton distributions which quite naturally lead
to rather unique expectations \cite{ref1} concerning the $P^2$ dependence
of the photonic parton distributions. Similar approaches have been
presented recently \cite{ref2,ref3}. Studies of parton distributions
of virtual $(P^2\neq 0)$ photons, $f^{\gamma (P^2)}(x,\mu_F^2)$ with
some properly chosen factorization scale $\mu_F$, may thus provide an
important test of the underlying ideas. In the present paper we present
predictions for production processes at the DESY-HERA $ep$ collider
which are highly sensitive to the 'resolved' components $f^{\gamma (P^2)}$.
In particular the predicted rates for the 'resolved' production 
rates in these
processes are presented in appropriate kinematical ranges suitable for
testing the recently proposed [1-3] $f^{\gamma (P^2)}$.
Detailed experimental tests of such predictions will obviously
elucidate the so
far unanswered question as to when a deep inelastic $ep$ process is
dominated by the usual 'direct' $\gamma^*\equiv\gamma (P^2)$ induced cross
sections, {\underline{not}} contaminated by the so far poorly known
'resolved' $f^{\gamma (P^2)}(x,\mu_F^2)$ contributions.

To introduce our notations and conventions we begin in Sec.\ 2 with a short
presentation of the proposed model \cite{ref1}. In Sec.\ 3 and 4 we present
quantitative results for the electroproduction of heavy quarks and
(single inclusive) jets, respectively, with detailed comparisons of the
individual direct and resolved contributions of virtual photons. Our
conclusions are drawn in Sec.\ 5. In the Appendix we present a LO-QCD
parametrization of our predicted distributions \cite{ref1}
which should be sufficient
and useful for any forthcoming experimental or theoretical investigation
as well as for eventual Monte Carlo analyses.
%
%
\section{The Model}
The basic ingredient of our model for the parton content of virtual
transverse photons $\gamma (P^2)$ is that it should be smooth in $P^2$
at $P^2=0$ where previous results \cite{ref4} for the real photon should
hold, which are given at the low (dynamical) input scale $Q^2=\mu^2$
by
\begin{equation}
f^{\gamma}(x,\mu^2) =\kappa\; (4 \pi \alpha/f_{\rho}^2)\; f^{\pi}(x,\mu^2)
\end{equation}
where $f=q,\;\bar{q},\;g$ and with $\kappa,\;f_{\rho}^2,\;\mu^2$ and
$f^{\pi}(x,\mu^2)$ specified in \cite{ref4} and \cite{ref5}, respectively.
This is expected to hold for the leading order (LO) distributions as well
as for the next-to-leading order (NLO) real photon $[\gamma\equiv
\gamma (P^2=0)]$ distributions in the ${\rm{DIS}}_{\gamma}$
factorization scheme \cite{ref6} which are related to the
${\overline{\rm{MS}}}$ 
factorization scheme distributions via \cite{ref4,ref6}
\begin{equation}
f_{{\rm{DIS}}_{\gamma}}^{\gamma} (x,Q^2) =
f_{\overline{{\rm{MS}}}}^{\gamma} (x,Q^2) + \delta f^{\gamma} (x)
\end{equation}
with
\begin{eqnarray}
\nonumber
\delta q^{\gamma} (x) = \delta \bar{q}^{\gamma}(x) & = &
3 e_q^2 \frac{\alpha}{2 \pi} \left\{ \left[x^2+(1-x)^2 \right] \ln
\frac{1-x}{x} -1 +8x -8x^2 \right\} \\
\delta g^{\gamma}(x) & = & 0\;\;\;\;.
\end{eqnarray}
This, together with a VMD inspired $P^2$ dependence factor yields the
following boundary conditions for $f^{\gamma (P^2)}(x,Q^2)$ at
$Q^2=\tilde{P}^2$:
\begin{equation}
\nonumber
f^{\gamma (P^2)}(x,\tilde{P}^2) = \eta(P^2)
f_{non-pert}^{\gamma (P^2)}
(x,\tilde{P}^2) + \left[ 1- \eta(P^2)\right]
f_{pert}^{\gamma (P^2)}(x,\tilde{P}^2)
\end{equation}
with $\tilde{P}^2=\max (P^2,\mu^2)$ and
%
$ \eta(P^2) = (1 +P^2/m_{\rho}^2)^{-2} $
%
where $m_{\rho}^2=0.59{\rm{GeV}}^2$ refers to some effective mass
in the vector-meson propagator; furthermore
\begin{equation}
f_{non-pert}^{\gamma (P^2)}(x,\tilde{P}^2) =
\kappa\; (4\pi \alpha/f_{\rho}^2)
\left\{ \begin{array}{ccc}
f^{\pi}(x,P^2)& , &P^2>\mu^2 \\
& & \\
f^{\pi}(x,\mu^2) &,& 0\le P^2 \le \mu^2
\end{array} \right.
\end{equation}
and
\begin{eqnarray}
\nonumber
q_{pert}^{\gamma (P^2)}(x,\tilde{P}^2) &=& \bar{q}_{pert}^{\gamma (P^2)}
(x,\tilde{P^2}) = \left\{\begin{array}{l}
0\;,\;\;{\rm{LO}}\\
3 e_q^2 \frac{\alpha}{2\pi} \left[\left(x^2+(1-x)^2\right)\ln\frac{1}{x^2}-2
+6x-6x^2\right]\;,\;\;{\rm{NLO}}\end{array} \right.\\
g_{pert}^{\gamma (P^2)}(x,\tilde{P}^2)&=& 0
\end{eqnarray}
where the NLO boundary conditions are again specified in the
${\rm{DIS}}_{\gamma}$ 
factorization scheme and $\mu_{LO}^2=0.25\,{\rm{GeV}}^2$
and $\mu_{NLO}^2=0.3\,{\rm{GeV}}^2$ refer to the LO and NLO input scales
for the parton distributions of the real photon. The boundary conditions
(2.5) and (2.6) apply both in LO and NLO. The resulting LO distributions at
$Q^2\ge \tilde{P}^2$ are presented in the Appendix and are applicable
at $Q^2\gg P^2$ where higher twist $(P^2/Q^2)^n$ contributions are likely
to be negligible and transverse photon contributions dominate.
%
%
\section{Heavy Quark Electroproduction}
\setcounter{equation}{0}
The cross section for heavy quark $(h=c,\,b,\,\ldots)$
electroproduction is given by
\begin{equation}
\frac{d\sigma_{ep}^h}{dP^2dy}= \frac{2\pi\alpha^2}{(P^2)^2y}
\left[ (1+(1-y)^2) F_2^h(x,P^2)- y^2 F_L^h(x,P^2) \right]
\end{equation}
with $x=P^2/sy$ and where the 'direct' contribution to $F_{2,L}^h$ arise in
LO from $\gamma (P^2) g\rightarrow h\bar{h}$,
\begin{equation}
\frac{1}{x} F_{2,L}^h(x,P^2) = 2 e_h^2\; \frac{\alpha_s(\mu_F^2)}{2\pi}
\int_{ax}^1 \frac{dx'}{x'}\; C_{2,L}\left( \frac{x}{x'},\rho \right)
g(x',\mu_F^2)
\end{equation}
with
\begin{eqnarray}
\nonumber
C_2(z,\rho) &=& \frac{1}{2}\;\Bigg\{ 
\left[z^2+(1-z)^2+z(1-3z)\rho -\frac{1}{2}
z^2 \rho^2 \right] \ln \frac{1+\beta}{1-\beta} \\
&+& \beta \;\left[-1+8z(1-z)-z(1-z) \rho\right] \Bigg\} \\
C_L(z,\rho) &=& -z^2 \rho \;\ln \frac{1+\beta}{1-\beta}+2\beta\; z(1-z)
\end{eqnarray}
where $\rho\equiv 4 m_h^2/P^2,\;a=1+\rho,\; \beta^2=1-\rho\; z(1-z)^{-1}$
and the factorization scale $\mu_F$ should be preferrably chosen at
$\mu_F^2=4 m_h^2$, irrespective of $P^2$, which results in a satisfactory
perturbative stability when compared with the appropriate NLO
contribution \cite{ref7}. The description of the above 'direct' heavy quark
electroproduction in terms of $\sigma_{\gamma p}^h$ for the process
$\gamma p\rightarrow h\bar{h}$ with $\gamma$ denoting an effective,
transversely polarized, {\underline{massless}} photon, $\gamma \equiv
\gamma(P^2=0)$, yields the approximation \cite{ref8}
\begin{equation}
\frac{d\sigma_{ep}^h}{dP^2dy} \simeq \frac{\alpha}{2\pi} \frac{1}{P^2}
\frac{1+(1-y)^2}{y} \sigma_{\gamma p}^h (s y)
\end{equation}
where
\begin{equation}
\sigma_{\gamma p}^h(s y) = \frac{1}{sy} \int_{4 m_h^2}^{sy} d\hat{s}\;
g\left(\frac{\hat{s}}{sy},\mu_F^2\right) \hat{\sigma}_{\gamma g}^h
(\hat{s})\;\;\;,
\end{equation}
with the cross section for the 'direct' subprocess $\gamma g \rightarrow
h\bar{h}$ being given by
\begin{equation}
\hat{\sigma}^h_{\gamma g}(\hat{s}) = e_h^2\; \frac{2\pi \alpha
\alpha_s(\mu_F^2)}{\hat{s}} \left[ \frac{1}{2} (3-\hat{\beta}^4)
\ln \frac{1+\hat{\beta}}{1-\hat{\beta}}-\hat{\beta}(2-\hat{\beta}^2)
\right]
\end{equation}
where $\hat{\beta}^2=1-4m_h^2/\hat{s}$.
The validity of this approximation determines the allowed range in $P^2$
where the virtual photon concept, in the sense discussed in [1-3],
is meaningful. We recall that our expectations concerning this issue
are that as long as $P^2\ll\mu_F^2\simeq 4 m_h^2$ is fulfilled, the
effective virtual photon concept is useful and the study of its partonic
content applicable to investigations concerning the so called 'resolved
photon' contributions. In fact we find that as long as $P^2/\mu_F^2\,
$\raisebox{-1mm}{${\stackrel{\textstyle <}{\sim}}$}$\,10^{-1}$ the
effective virtual photon approximation, Eq.(3.5), reproduces the results
of the exact calculation, Eq.(3.1), at a level of up to about 10$\%$
accuracy. This is illustrated in Tables 1 and 2 for charm and bottom
production, respectively, where we used $\mu_F^2=4m_h^2$ together with
$m_c=1.5\,{\rm{GeV}}$ and $m_b=4.5\,{\rm{GeV}}$. [Note that, for a given
$P^2$, the validity of the approximation (3.5) is almost independent
of $y$ due to the smallness of $F_L^h$ in (3.1)]. We thus expect that
in general the effective photon concept is reasonably applicable
whenever $P^2\,$\raisebox{-1mm}{${\stackrel{\textstyle <}{\sim}}$}$\,
10^{-1} \mu_F^2$ and that its resolved parton content is discernible
provided it contributes significantly more than 10$\%$ to the involved
production rate. This latter requirement is necessary also due to
additional uncertainties related to different choices of the
factorization scale (e.g., $\mu_F^2=m_h^2$) and to the size of NLO
corrections \cite{ref1,ref7}. As shown in Figs.\ 1 and 2, these
conditions are almost {\underline{never}} 
satisfied for heavy quark $(c,\;b)$
electroproduction where the resolved contribution, to be added to the
direct $\sigma_{\gamma p}^h$ in (3.5), is given by \cite{ref9}
\begin{equation}
\sigma_{\gamma p,res}^h(s_{\gamma^*p}) = \sum_{f^{\gamma}}
\sum_f \int dx_{\gamma} dx \;f^{\gamma (P^2)}(x_{\gamma},\mu_F^2)
f(x,\mu_F^2) \hat{\sigma}^{f^{\gamma}f}(x_{\gamma}x s_{\gamma^*p},
m_h^2,\mu_F^2)
\end{equation}
with $s_{\gamma^*p}=ys-P^2$. The dominant LO hadronic $2\rightarrow 2$
(as well as NLO $2\rightarrow 3$) subprocess cross sections
$\hat{\sigma}^{f^{\gamma}f}$ for $h\bar{h}$ production, i.e.,
$f^{\gamma}f\rightarrow h\bar{h}$ etc.\ with $f=q,\;\bar{q},\;g$, are
well known \cite{ref9}. Due to the smallness of the hadronic 'resolved'
contributions for $P^2\equiv Q^2\neq 0$ in Figs.\ 1 and 2,
electroproduction of heavy quarks appears to be unsuitable for studies
of the resolved parton content of the virtual photon at HERA -
{\em{unless}} one
can tag experimentally on the resolved contribution via the
{\em{hadronic photon remnants}}.
The situation here is not too different from that for real
photons \cite{ref9}, $P^2=0$, and improves for ${\rm{LEP}}*{\rm{LHC}}\;
(\sqrt{s}\simeq 1265\,{\rm{GeV}})$. For $h=b$ (Fig.\ 2) a $P^2$ dependence
of the resolved component may be discernible for $P^2<1 {\rm{GeV}}^2$ where
the effective photon description is accurate at a level of 1$\%$ (Table 2)
and the resolved component's contribution is about or exceeds 10$\%$.

On the other hand, the smallness of the resolved contribution of virtual
($P^2\equiv Q^2>0$) photons to heavy quark production implies that $ep$
scattering at HERA allows for reliable predictions of heavy quark production
rates uncontaminated by the resolved contributions. Finally it should be
noted that photoproduction ($P^2=0)$ of charm, $\gamma p\rightarrow
c \bar{c} X$, has already been measured at HERA \cite{ref11,ref12} and
first preliminary measurements for electroproduction $(P^2>0)$ of charm,
$ep\rightarrow e\,c\bar{c}\,X$, i.e, for $F_2^c(x,P^2\equiv Q^2)$ have
recently appeared \cite{ref13}.
%
\section{Jet Electroproduction}
\setcounter{equation}{0}
Turning now to high-$E_T$ single-inclusive jet production,
$ep\rightarrow jet+X$, we shall study the $P^2$ dependence of the
resolved photon contribution in the range
$P^2\,$\raisebox{-1mm}{${\stackrel{\textstyle <}{\sim}}$}$\,10^{-1}\mu_F^2$
indicated in the previous section as suitable for an effective photon
description. It now turns out that in large kinematical regions the
resolved photon contribution to the jet production rate is indeed
significant and thus allows for a reliable study of its $P^2$ dependence.
For definiteness we shall study $d\sigma^{jet}_{ep}/dP^2dE_Tdyd\eta_{lab}$
at the HERA $ep$ collider with realistic values chosen for $E_T,\;y$ and
$\eta_{lab}$, where $d\sigma_{ep}^{jet}$ replaces $d\sigma_{ep}^h$ in
(3.5) and similarly $\sigma_{\gamma p}^h$ 
is replaced by the direct and resolved
contributions
\begin{eqnarray}
\nonumber
d\sigma_{\gamma p}^{jet} &=& d\sigma_{dir}^{jet}+d\sigma_{res}^{jet}\\
\nonumber
&=& \sum_{f=q,\bar{q},g} \int dx\; f(x,\mu_F^2) d\hat{\sigma}_{\gamma f}
^{jet} \\
&+& \sum_{f^{\gamma}}\sum_f \int dx_{\gamma} dx \; f^{\gamma (P^2)}
(x_{\gamma},\mu_F^2) f(x,\mu_F^2) d\hat{\sigma}_{f^{\gamma}f}^{jet}\;\;\,.
\end{eqnarray}
The LO $\gamma f\rightarrow q\bar{q}$ and hadronic $2\rightarrow 2$
subprocess cross sections $d\hat{\sigma}^{jet}_{\gamma f}$ and
$d\hat{\sigma}^{jet}_{f^{\gamma}f}$, respectively, can be found, for
example, in \cite{ref14}. The charmed quark jet's contribution is
included via $\gamma g\rightarrow c\bar{c}$ in the 'direct' channel and in
the 'resolved' channel via $g^{\gamma (P^2)} g\rightarrow c\bar{c}$,
etc., with the relevant unintegrated matrix elements being given in
\cite{ref15} and \cite{ref16}, respectively. The sensitivity to the
actual value of $m_c$ is small due to $m_c^2/E_T^2\ll 1$ for
typically chosen values of $E_T$. The chosen $E_T$ of the jet for
realistic experiments should not be too small for several obvious reasons:
(i) The jet should be clearly separable from possible beam contaminations,
(ii) $P^2\,$\raisebox{-1mm}{${\stackrel{\textstyle <}{\sim}}$}$\,10^{-1}
E_T^2$ should be fulfilled over a wide range of $P^2$ and (iii)
perturbative calculations of its production rate should be reliable.
On the other hand $E_T$ should not be too large in order that (i)
a reasonable signal is still available and (ii) the resolved contribution
is still significant.
The choice $E_T\,$\raisebox{-1mm}{${\stackrel{\textstyle >}{\sim}}$}$\,
5-7\,{\rm{GeV}}$ and 
consequently $\mu_F\sim E_T$, as also employed by recent
experimental analyses of photoproduced single-jet \cite{ref17}
and dijet \cite{ref18} events, seems to meet all these requirements.
As for $y$ we will choose $y\simeq 0.5$ as representative for the
realistic range $0.2\,$\raisebox{-1mm}{${\stackrel{\textstyle <}{\sim}}$}$\,
y\,$\raisebox{-1mm}{${\stackrel{\textstyle <}{\sim}}$}$\,0.8$ containing
a sufficient flux of energetic photons. Finally, the pseudorapidity
$\eta_{lab}$ in the HERA lab-frame should be chosen so as to guarantee
a significant contribution of the resolved partons to $d\sigma^{jet}$
over a wide range of $P^2$. This happens for \cite{ref19}
$-1\,$\raisebox{-1mm}{${\stackrel{\textstyle <}{\sim}}$}$\,\eta_{lab}\,
$\raisebox{-1mm}{${\stackrel{\textstyle <}{\sim}}$}$\,2.5$.
We shall evaluate the resulting relevant ratio [cf. Eq.(4.1)]
\begin{equation}
\left(\frac{res}{dir}\right)^{jet} \equiv d\sigma_{res}^{jet}
(P^2;\;E_T,\,y,\;\eta_{lab})/ d\sigma_{dir}^{jet}
(P^2;\;E_T,\,y,\;\eta_{lab})
\end{equation}
for the above specified values of $E_T$, $y$ and $\eta_{lab}$.
This ratio will be analysed in LO since NLO corrections are expected
to be less significant in the above ratio. Moreover the NLO/LO stability
for photoproduction $(P^2=0)$ of 
jets has been demonstrated \cite{ref20,ref21} as
well as the stability with respect to the choice of the renormalization
and factorization scale $\mu_F\simeq E_T$. Most of our quantitative
results are based on using in Eq.(4.1) the LO parton distributions of the
proton in \cite{ref10} and of the virtual photon in \cite{ref1},
unless stated otherwise.

The result of our numerical calculations are shown in Figs.\ 3-5 as a
function of $P^2$ for various fixed values of $E_T$ and $\eta_{lab}$ at
a typical average $y\simeq 0.5$. These demonstrate clearly the
feasibility of investigating the $P^2$ dependent parton distributions
of the virtual photon at HERA, in particular of course in the forward
(proton) beam direction $\eta_{lab}>0$ (i.e.\ small $x_{\gamma}$), up to
$P^2\simeq 5\,{\rm{GeV}}^2$, i.e., up to $P^2/E_T^2\simeq 10^{-1}$,
where the effective virtual photon concept is still meaningful. It is
interesting to note that even for $\eta_{lab}\le 0$ (large(r) $x_{\gamma}$),
the resolved component is still comparable to the direct one, i.e.,
$res/dir\,$\raisebox{-1mm}{${\stackrel{\textstyle <}{\sim}}$}$\,3$ as
shown in Fig.\ 4. This implies that jet production in $ep$ collisions
by virtual $(P^2\equiv Q^2\neq 0$) photons at HERA energies rarely proceeds
via the 'direct' $\gamma^*$-contribution alone, since the resolved
components of the virtual photon remain effective for
$-1\,$\raisebox{-1mm}{${\stackrel{\textstyle <}{\sim}}$}$\,\eta_{lab}\,
$\raisebox{-1mm}{${\stackrel{\textstyle <}{\sim}}$}$\,2.5$ at the relevant
$E_T$ and $P^2$. It should be mentioned 
that preliminary results \cite{ref22}
on the integrated event-rate for dijet production in $ep$ collisions
with virtual $(0.1<P^2<0.55\,{\rm{GeV}}^2)$ photons have recently
demonstrated the relevance of their resolved components.

Finally, in Fig.\ 6 we compare our results with those obtained
utilizing the parton distributions of the virtual photon in \cite{ref2}:
SaS 1D and SaS 2D refer to rather different virtual photon parton
densities, corresponding to evolution input scales
$Q_0^2=0.36\,{\rm{GeV}}^2$ and $4\,{\rm{GeV}}^2$, respectively, from where
appropriately modified VMD inputs are evolved. The predicted $P^2$
dependence of $'res/dir'$ in Fig.\ 6 is not too different from the one
based on the GRS parton distributions \cite{ref1} of the virtual
photon. Similar results hold for other values of $E_T$ and $\eta_{lab}$
considered previously.
%
\section{Summary and Conclusions}
Parton densities of virtual $(P^2\neq 0)$ photons can be rather uniquely
calculated in LO and NLO QCD which, moreover, extrapolate
smoothly to $P^2=0$, i.e., to the hadronic 'resolved' structure functions
of a real photon \cite{ref1}. The virtual photon structure function
$F_2^{\gamma (P^2)}(x,Q^2)$ can be measured via the subprocess
$\gamma^*(Q^2) \gamma(P^2)\rightarrow X$ in $e^+ e^-\rightarrow
e^+e^- X$ at LEP2 \cite{ref23} or in tagged $ep\rightarrow e X$
collisions at DESY-HERA via the subprocess $\gamma (P^2)p\rightarrow X$
where the parton densities $f^{\gamma (P^2)}(x,\mu_F^2)$,
$f=q,\;\bar{q},\;g$ and $P^2\equiv Q^2$, describe the 'resolved' hadronic
component of the virtual photon $\gamma^*\equiv \gamma(P^2)$.

Here we have shown that the predicted [1-2] $P^2$ dependence of the parton
distributions $f^{\gamma (P^2)}$ inside the virtual photon may be tested
at the DESY-HERA $ep$ collider via tagged measurements of single jet or
$b\bar{b}$ pair production. In particular the observation of single
jets with transverse energy $E_T^{jet}\simeq 5-7\,{\rm{GeV}}$ and
pseudorapidity $\eta_{lab}>0$ will provide for a reliable determination
of $f^{\gamma (P^2)}(x_{\gamma},E_T^2)$ at
$P^2\,$\raisebox{-1mm}{${\stackrel{\textstyle <}{\sim}}$}$\,5\,{\rm{GeV}}^2$.
For $\eta_{lab}\le 0$, where larger values of $x_{\gamma}$ are probed,
the contributions due to the 'resolved' components
$f^{\gamma (P^2)} (x_{\gamma},E_T^2)$ are still comparable to the 'direct'
contribution from the unresolved $\gamma(P^2)$.

A determination of $g^{\gamma (P^2)}(x_{\gamma}, 4 m_b^2)$ is shown to be
feasible at $P^2\,$\raisebox{-1mm}{${\stackrel{\textstyle <}{\sim}}$}$\,
1{\rm{GeV}}^2$ where the resolved contribution to
$d\sigma (ep\rightarrow e b\bar{b} X)$ amounts to about 10$\%$. As a
byproduct we note that all the delineated kinematical regions where
the resolved contributions due to $f^{\gamma (P^2)}$ are negligible, such
as for example in $c\bar{c}$ production, are particularly suitable for a
reliable determination of parton distributions inside the
{\underline{proton}}: Most important is the possibility of fixing
$g(x,4m_c^2)$ via $ep\rightarrow e\,c\bar{c}\,X$ at all conceivable
values of $P^2$. Furthermore, $g^{\gamma (P^2)}(x_{\gamma}, 4m_c^2)$ can
be determined even in the aforementioned $ep\rightarrow e\,c\bar{c}\,X$
process {\em{provided}} the
{\em{hadronic remnants}} of the photon can be detected
experimentally. A similar remark obviously pertains also for jet events,
in particular for those with $\eta_{lab}<0$ relevant for the
larger-$x_{\gamma}$ region.

Finally it should be stressed again that, according to our analysis in
Sec.\ 3, the concept of a resolved virtual photon is only meaningful
for virtualities constrained by
$P^2\,$\raisebox{-1mm}{${\stackrel{\textstyle <}{\sim}}$}$\,10^{-1}
\mu_F^2$ with $\mu_F^2$ denoting the typical momentum scale of the
underlying hard processes.
%
\section*{Acknowledgement}
This work has been supported in part by the
'Bundesministerium f\"ur Bildung, Wissenschaft, Forschung und Technologie',
Bonn.
%
%
\section*{Appendix}
\setcounter{equation}{0}
\renewcommand{\theequation}{A\arabic{equation}}
In order to obtain parametrizations 
of our radiative (dynamical) LO predictions
\cite{ref1} for the parton densities $f^{\gamma (P^2)}(x,Q^2)$ of the
virtual photon $\gamma (P^2)$, valid for $Q^2>\mu_{LO}^2=0.25\,{\rm{GeV}}^2$,
we write
\begin{equation}
xf^{\gamma (P^2)}(x,Q^2) = x f_{PL}^{\gamma (P^2)}(x,Q^2) +
x f_{HAD}^{\gamma (P^2)}(x,Q^2)\;\;\;.
\end{equation}
The pointlike (PL) contributions \cite{ref1} are parametrized as
$(\alpha=1/137)$
\begin{equation}
xf_{PL}^{\gamma (P^2)}(x,Q^2)/\alpha = \left[ s\, x^a \left(
A+B\sqrt{x}+C x^b\right)+s^{\alpha} e^{-E+\sqrt{E'\;s^{\beta} \ln \frac{1}
{x}}}\right]\;(1-x)^D
\end{equation}
where
\begin{equation}
s=\ln\frac{\ln \left(Q^2/(0.232\,{\rm{GeV}}\right)^2)}
{\ln \left(\tilde{P}^2/(0.232\,{\rm{GeV}})^2\right)}
\end{equation}
with $\tilde{P}^2=\max (P^2,\,\mu_{LO}^2)$. The $Q^2$ dependence of the
quantities $a,\,b,\,A,\,B,\,\ldots$ is described as usual in terms of
power series in $s$, as defined in Table 3, with coefficients
$F,\,G,\,H$ which in turn are, according to Table 3, expanded as, e.g.\
\begin{equation}
G=g_1+g_2\,l_1(P^2)+g_3\,l_2(P^2)
\end{equation}
in order to describe the $P^2$ dependence in (A2), with
\begin{equation}
l_1(P^2)=\ln^2(\tilde{P}^2/\mu^2_{LO})\;\;\;,\;\;\;
l_2(P^2)=\ln\left[\tilde{P}^2/\mu_{LO}^2+\ln(\tilde{P}^2/\mu^2_{LO})
\right]\;\;\;.
\end{equation}
All the required constants are tabulated in Table 3 for the pointlike
$u=\bar{u}$, $d=\bar{d}=s=\bar{s}$ and $g$ densities (A2).

The hadronic (HAD) contributions to (A1), being generated from the
VMD-like boundary conditions \cite{ref1}, are parametrized as
\begin{equation}
xf_{HAD}^{\gamma (P^2)}(x,Q^2)/\alpha = \eta (P^2)\,\left[ x^a \left(
A+B\sqrt{x}+C x^b\right)+s^{\alpha} e^{-E+\sqrt{E'\;s^{\beta} \ln \frac{1}
{x}}}\right]\;(1-x)^D
\end{equation}
for the hadronic $u=\bar{u}=d=\bar{d}$ and $g$ densities, and as
\begin{equation}
xs_{HAD}^{\gamma (P^2)}(x,Q^2)/\alpha = \eta (P^2)\,\left[s\, x^a \left(
A+B\sqrt{x}+C x^b\right)+s^{\alpha} e^{-E+\sqrt{E'\;s^{\beta} \ln \frac{1}
{x}}}\right]\;(1-x)^D
\end{equation}
for the $s=\bar{s}$ density, with $\eta (P^2)=(1+P^2/m_{\rho}^2)^{-2}$ and
$m_{\rho}^2=0.59\,{\rm{GeV}}^2$. The $Q^2$ dependence of (A6) and (A7) is
described by expanding again $a,\,b,\,A,\,B,\,\ldots$ in terms of
\begin{equation}
s=\ln\frac{\ln \left(Q^2/(0.232\,{\rm{GeV}}\right)^2)}
{\ln \left(\mu^2_{LO}/(0.232\,{\rm{GeV}})^2\right)}
\end{equation}
as described in Table 4.

All above parametrizations are valid for
\begin{eqnarray}
\nonumber
& P^2 &\!\!\!\! \le 10\,{\rm{GeV}}^2\;,\;\;\; 10^{-4}\le x\le 1\\
0.6\le\!\!\!\! & Q^2 & \!\!\!\!\le 5\times 10^4 \, 
{\rm{GeV}}^2\;\;\;{\rm{\underline{and}}}\;\;\;
Q^2 \gtrsim 5 P^2
\end{eqnarray}
and are obtainable as a {\sc{Fortran}} package  
via electronic mail from
strat@hal1.physik.uni-dortmund.de      
%

\newpage
%
%
%
\section*{Tables}
\begin{description}
\item[Table 1.] Cross sections for direct electroproduction of
charm ($m_c=1.5\,{\rm{GeV}}$) at HERA energies $\sqrt{s}=298\,{\rm{GeV}}$
for various photon virtualities $P^2$ and energy fractions $y$. The
exact results refer to Eqs.\ (3.1)-(3.4) and the approximate ones to
Eqs.\ (3.5)-(3.7), with the percental differences being shown in the
last column.
\end{description}
\begin{center}
\begin{tabular}{c|c|c|c|c|}
$P^2/{\rm{GeV}}^2$ & $y$ & EXACT (nb) & APPROX (nb) & $\%$ \\ \hline
{} & 0.1 & 369.2 & 378.5 & 2.52 \\ \cline{2-5}
0.2 & 0.5 & 108.4 & 111.1 & 2.49 \\ \cline{2-5}
& 0.7 & 77.98 & 80.02 & 2.62 \\ \hline
& 0.1 & 142.4 & 151.4 & 6.32 \\ \cline{2-5}
0.5 & 0.5 & 41.80 & 44.43 & 6.29 \\ \cline{2-5}
& 0.7 & 30.02 & 32.01 & 6.63 \\ \hline
& 0.1 & 67.09 & 75.70 & 12.8 \\ \cline{2-5}
1.0 & 0.5 & 19.71 & 22.22 & 12.7 \\ \cline{2-5}
& 0.7 & 14.12 & 16.00 & 13.3 \\ \hline
\end{tabular}
\end{center}
%
%
\begin{description}
\item[Table 2.]  As in Table 1 but for bottom production
($m_b=4.5\,{\rm{GeV}}$).
\end{description}
\begin{center}
\begin{tabular}{c|c|c|c|c|}
$P^2/{\rm{GeV}}^2$ & $y$ & EXACT (nb) & APPROX (nb) & $\%$ \\ \hline
{}& 0.1 & 0.962 & 0.970 & 0.83 \\ \cline{2-5}
0.5 & 0.5 & 0.404 & 0.407 & 0.74 \\ \cline{2-5}
& 0.7 & 0.308 & 0.311 & 0.97 \\ \hline
& 0.1 & 0.234 & 0.243 & 3.85 \\ \cline{2-5}
2.0 & 0.5 & 0.099 & 0.102 & 3.03 \\ \cline{2-5}
& 0.7 & 0.075 & 0.078 & 4.00 \\ \hline
& 0.1 & 0.089 & 0.097 & 8.99 \\ \cline{2-5}
5.0 & 0.5 & 0.038 & 0.041 & 7.89 \\ \cline{2-5}
& 0.7 & 0.029 & 0.031 & 6.90 \\ \hline
& 0.1 & 0.041 & 0.0485 & 18.29 \\ \cline{2-5}
10.0 & 0.5 & 0.0175 & 0.0204 & 16.57 \\ \cline{2-5}
& 0.7 & 0.0133 & 0.0155 & 16.54 \\ \hline
\end{tabular}
\end{center}
\newpage
\begin{description}
\item[Table 3] Expansion parameters for the parametrizations of the
pointlike (PL) contribution to (A1), according to Eqs.\ (A2)-(A5).
\end{description}
\hbox to \textwidth {\hss
\footnotesize
\begin{tabular}{c||c|c|c|c|c|c|c|c|c|}
& \multicolumn{9}{c}{$u$} \\ \cline{2-10}
& \multicolumn{3}{c}{$F=f_1+f_2\,l_1(P^2)+f_3\,l_2(P^2)$}\vline&
  \multicolumn{3}{c}{$G=g_1+g_2\,l_1(P^2)+g_3\,l_2(P^2)$}\vline&
  \multicolumn{3}{c}{$H=h_1+h_2\,l_1(P^2)+h_3\,l_2(P^2)$}\vline
 \\ \cline{2-10}
&$f_1$&$f_2$&$f_3$&$g_1$&$g_2$&$g_3$&$h_1$&$h_2$&$h_3$ \\ \hline
$\alpha=F$&1.551&-0.139&0.783&-&-&-&-&-&- \\ \hline
$\beta=F$&0.105&0.132&0.087&-&-&-&-&-&- \\ \hline
$a=F+G\cdot s$&1.089&0.003&-0.0134&-0.172&0.009&-0.017&-&-&-\\ \hline
$b=F+G\,\sqrt{s}+H\,s^2$&
3.822&0.092&-0.516&-2.162&-0.085&0.439&0.533&0.013&0.108\\ \hline
$A=F+G\,\sqrt{s}$&0.377&-0.013&0.27&0.299&0.107&-0.097&-&-&-\\ \hline
$B=F+G\,s+H\,s^2$&
-0.467&-0.019&-0.272&-0.412&-0.167&0.138&0.2&0.076&0.026\\ \hline
$C=F+G\,s$&0.487&0.04&0.124&0.0766&0.064&-0.016&-&-&- \\ \hline
$D=F+G\,s$&0.119&0.011&-0.065&0.063&0.002&0.044&-&-&- \\ \hline
$E=F+G\,s$&7.605&0.057&-1.009&0.234&-0.057&0.622&-&-&- \\ \hline
$E'=F+G\,s$&-0.567&0.162&0.227&2.294&-0.172&-0.184&-&-&-\\ \hline\hline
& \multicolumn{9}{c}{$d=s$} \\ \cline{2-10}
& \multicolumn{3}{c}{$F=f_1+f_2\,l_1(P^2)+f_3\,l_2(P^2)$}\vline&
  \multicolumn{3}{c}{$G=g_1+g_2\,l_1(P^2)+g_3\,l_2(P^2)$}\vline&
  \multicolumn{3}{c}{$H=h_1+h_2\,l_1(P^2)+h_3\,l_2(P^2)$}\vline \\
  \cline{2-10}
&$f_1$&$f_2$&$f_3$&$g_1$&$g_2$&$g_3$&$h_1$&$h_2$&$h_3$ \\ \hline
$\alpha=F$&2.484&0.033&0.007&-&-&-&-&-&- \\ \hline
$\beta=F$&1.214&-0.0516&0.12&-&-&-&-&-&- \\ \hline
$a=F+G\,s$&1.088&0.001&-0.013&-0.1735&0.018&-0.028&-&-&- \\ \hline
$b=F+G\,\sqrt{s}+H\,s^2$&
4.293&0.102&-0.595&-2.802&-0.114&0.669&0.5975&0.022&0.001\\ \hline
$A=F+G\,s$&0.128&0.004&0.054&0.0337&0.025&-0.02&-&-&-\\ \hline
$B=F+G\,s+H\,s^2$&
-0.1193&-0.003&-0.0583&-0.0872&-0.041&0.035&0.0418&0.009&0.009\\ \hline
$C=F+G\,s$&0.127&0.007&0.032&0.0135&0.021&-0.009&-&-&- \\ \hline
$D=F+G\,s$&0.14&0.01&-0.06&0.0423&0.004&0.036&-&-&- \\ \hline
$E=F+G\,s$&6.946&-0.067&-0.39&0.814&0.06&0.033&-&-&-\\ \hline
$E'=F+G\,s$&1.531&-0.148&0.245&0.124&0.13&-0.171&-&-&- \\ \hline\hline
& \multicolumn{9}{c}{$g$} \\ \cline{2-10}
& \multicolumn{3}{c}{$F=f_1+f_2\,l_1(P^2)+f_3\,l_2(P^2)$} \vline&
  \multicolumn{3}{c}{$G=g_1+g_2\,l_1(P^2)+g_3\,l_2(P^2)$} \vline&
  \multicolumn{3}{c}{$H=h_1+h_2\,l_1(P^2)+h_3\,l_2(P^2)$} \vline \\
  \cline{2-10}
&$f_1$&$f_2$&$f_3$&$g_1$&$g_2$&$g_3$&$h_1$&$h_2$&$h_3$ \\ \hline
$\alpha=F$&1.682&0.025&-&-&-&-&-&-&- \\ \hline
$\beta=F$&1.1&-0.018&0.112&-&-&-&-&-&- \\ \hline
$a=F+G\,\sqrt{s}$&0.5888&-0.025&0.177&-0.4714&-0.022&0.024&-&-&- \\ \hline
$b=F+G\,s^2$&0.5362&0.001&-0.0104&0.0127&-&-&-&-&- \\ \hline
$A=F+G\,\sqrt{s}+H\,s^2$&
0.07825&0.0&0.053&0.05842&0.005&-0.058&0.08393&0.034&0.073 \\ \hline
$B=F+G\,s$&-2.438&-1.082&-1.666&0.03399&0.0&0.086&-&-&- \\ \hline
$C=F+G\,s^2$&2.348&1.08&1.63&-0.07182&-0.0256&-0.088&-&-&- \\ \hline
$D=F+G\,s+H\,s^2$&1.084&-&-&0.3098&-0.004&0.016&-0.07514&0.007&-0.012\\ \hline
$E=F+G\,s$&3.327&0.01&-0.673&1.1&0.126&-0.167&-&-&- \\ \hline
$E'=F+G\,s$&2.264&0.032&-0.227&0.2675&0.086&-0.159&-&-&- \\ \hline
\end{tabular}
\hss}
\normalsize
\setlength{\baselineskip}{0.75cm} 
\setlength{\parskip}{0.45cm}
\newpage
\begin{description}
\item[Table 4] Expansion parameters for the parametrizations of the hadronic
(HAD) contribution to (A1), according to Eqs.\ (A6)-(A8).
\end{description}
\hbox to \textwidth {\hss
\footnotesize
\begin{tabular}{c||c|c|c||c|c|c||c|c|c|}
& \multicolumn{3}{c||}{$u=d$} &
  \multicolumn{3}{c||}{$s$} &
  \multicolumn{3}{c|}{$g$} \\ \cline{2-10}
&$F$&$G$&$H$&$F$&$G$&$H$&$F$&$G$&$H$ \\ \hline
$\alpha=F$&0.756&-&-&0.902&-&-&0.364&-&-   \\ \hline
$\beta=F$&0.187&-&-&0.182&-&-&1.31&-&-     \\ \hline
$a=F+G\,\sqrt{s}+H\,s$&0.109&-&-0.163&0.271&-&-0.346&0.86&-0.254&-  \\ \hline
$b=F+G\,s+H\,s^2$&22.53&-21.02&5.608&17.1&-13.29&6.519&0.611&-&0.008 \\ \hline
$A=F+G\,\sqrt{s}+H\,s^2$&
0.002&0.004&-&0.017&-0.01&-&-0.843&2.248&-0.201 \\ \hline
$B=F+G\,s+H\,s^2$&
0.332&-0.008&-0.021&0.031&-0.0176&0.003&-0.097&-2.412& \\ \hline
$C=F+G\,s+H\,s^2$&0.054&-0.039&-&-0.011&0.0065&-&1.33&-&0.572 \\ \hline
$D=F+G\,s+H\,s^2$&0.381&0.572&-&1.243&0.804&-&0.44&1.233&0.009 \\ \hline
$E=F+G\,s$&4.774&1.436&-&4.709&1.499&-&0.954&1.862&- \\ \hline
$E'=F+G\,s$&-0.614&3.548&-&-0.48&3.401&-&3.791&-0.079&- \\ \hline
\end{tabular}
\hss}
\normalsize
\setlength{\baselineskip}{0.75cm}
\setlength{\parskip}{0.45cm}
\newpage
%
%
\section*{Figure Captions}
\begin{description}
\item[Fig.\ 1] The ratio of the 'resolved' [Eq.(3.8)] to the 'direct'
[Eq.(3.6)] contribution to charm production in LO-QCD for various fixed
values of $P^2$ (in ${\rm{GeV}}^2$ units), using $m_c=1.5\,{\rm{GeV}}$ and
$\mu_F^2=4 m_c^2$. The GRV 94 parton densities of the proton are taken from
\cite{ref10} and the ones of the virtual photon from \cite{ref1}.
For comparison the total $ep$ (dir.+res.) charm
production rate at HERA 
for $P^2=0$ is about $0.5\,{\mu}{\rm{b}}$ \cite{ref9} at
$\sqrt{s} \simeq 300\,{\rm{GeV}}$.
\item[Fig.\ 2] As in Fig.\ 1 but for bottom production, using
$m_b=4.5\,{\rm{GeV}}$ and $\mu_F^2=4 m_b^2$. The total $ep$ (dir.+res.)
bottom production rate at HERA for $P^2=0$ is about 6 nb \cite{ref9}
at $\sqrt{s}\simeq 300\,{\rm{GeV}}$.
\item[Fig.\ 3] The ratio of the 'resolved' to the 'direct' contribution,
defined in Eq.(4.2), for inclusive single jet production in LO-QCD in
$ep$ collisions at $\sqrt{s}=298 \,{\rm{GeV}}$ as a function of $P^2$.
The LO parton densities of the virtual photon are from \cite{ref1} and
the ones of the proton are taken from \cite{ref10}. Furthermore
$\mu_F^2=E_T^2$.
\item[Fig.\ 4] As in Fig.\ 3 but for $\eta_{lab}\le0$.
\item[Fig.\ 5] As in Fig.\ 3 but for different values of $E_T$ at fixed
$\eta_{lab}=1.5$.
\item[Fig.\ 6] Comparison of our results in Fig.\ 3 at $\eta_{lab}=1.5$
with the predictions due to two other representative sets \cite{ref2}
of parton densities of the virtual photon.
\end{description}
\newpage
\pagestyle{empty}

\vspace*{-0.0cm}
\hspace*{-1.4cm}
\epsfig{file=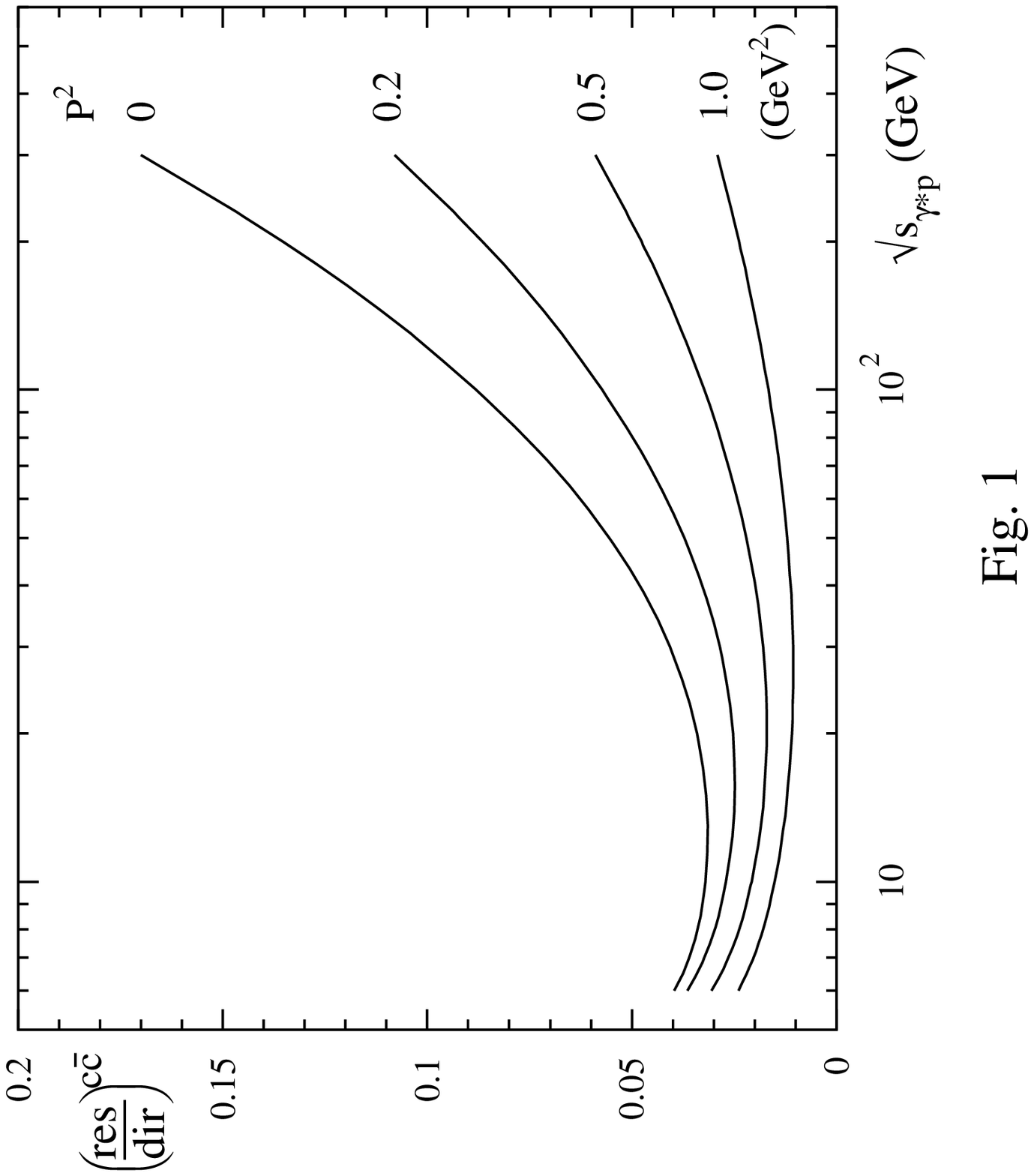,angle=90}
\newpage

\vspace*{-0.0cm}
\hspace*{-1.4cm}
\epsfig{file=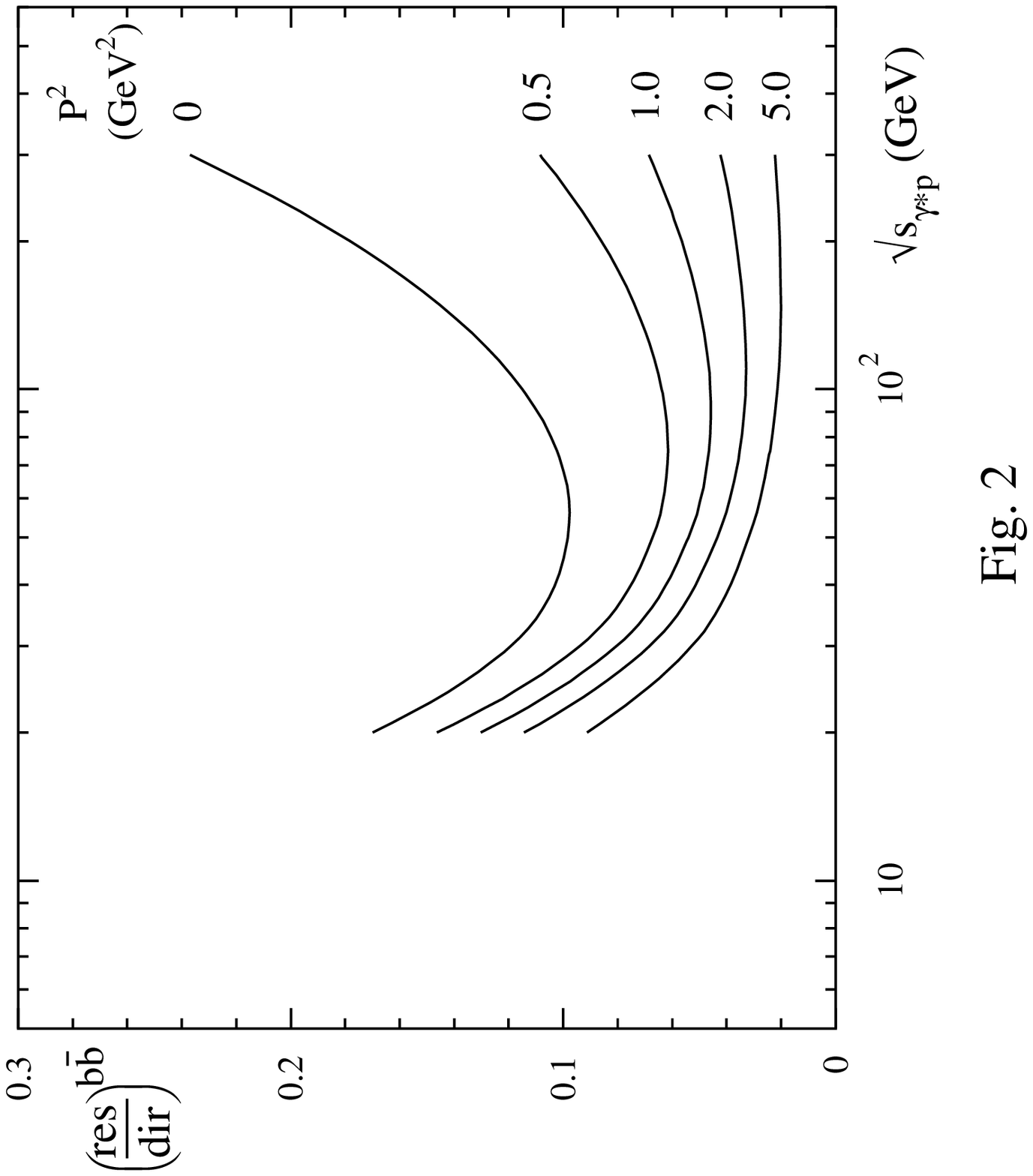,angle=90}
\newpage

\vspace*{-0.0cm}
\hspace*{-1.4cm}
\epsfig{file=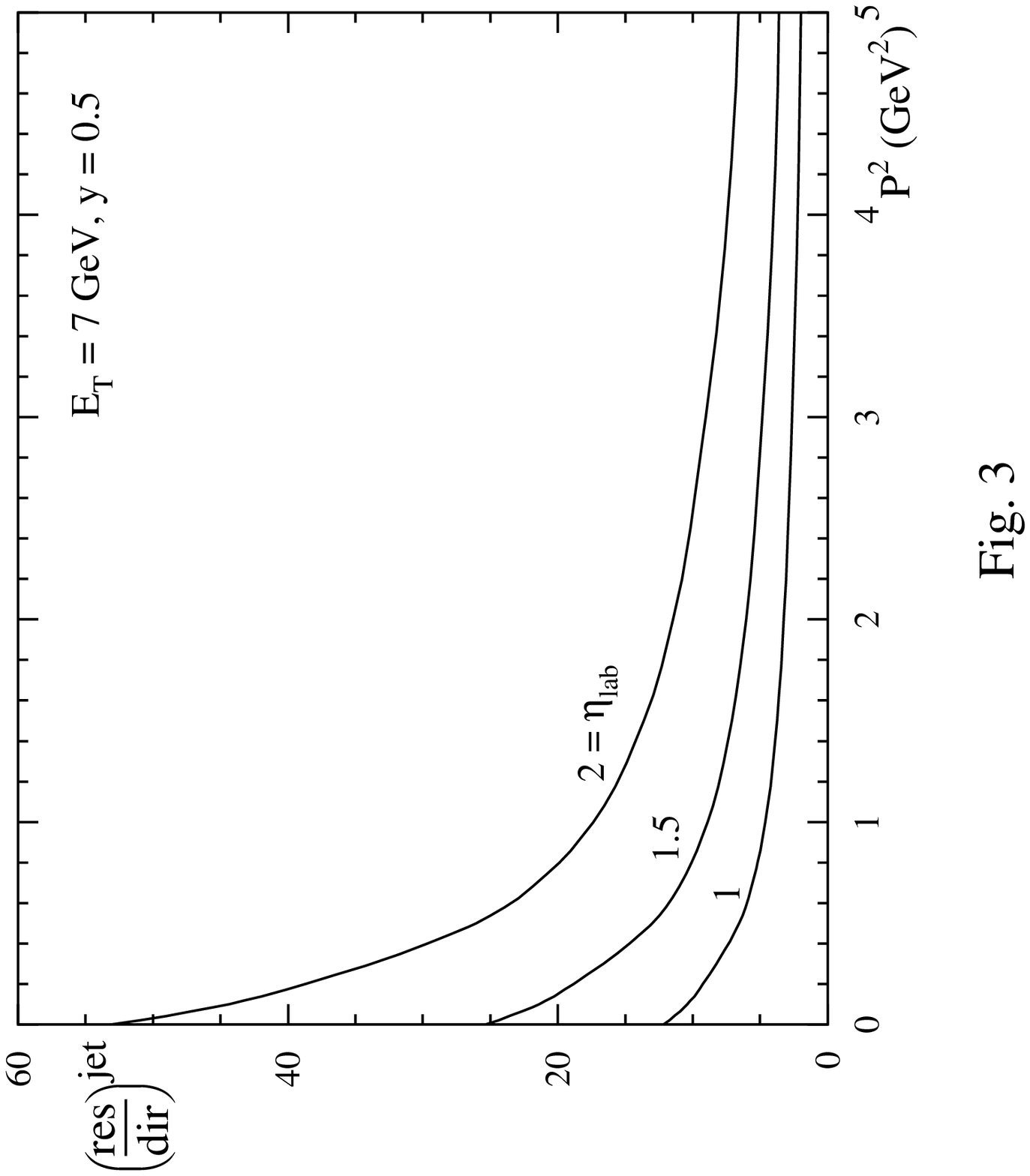,angle=90}
\newpage

\vspace*{-0.0cm}
\hspace*{-1.4cm}
\epsfig{file=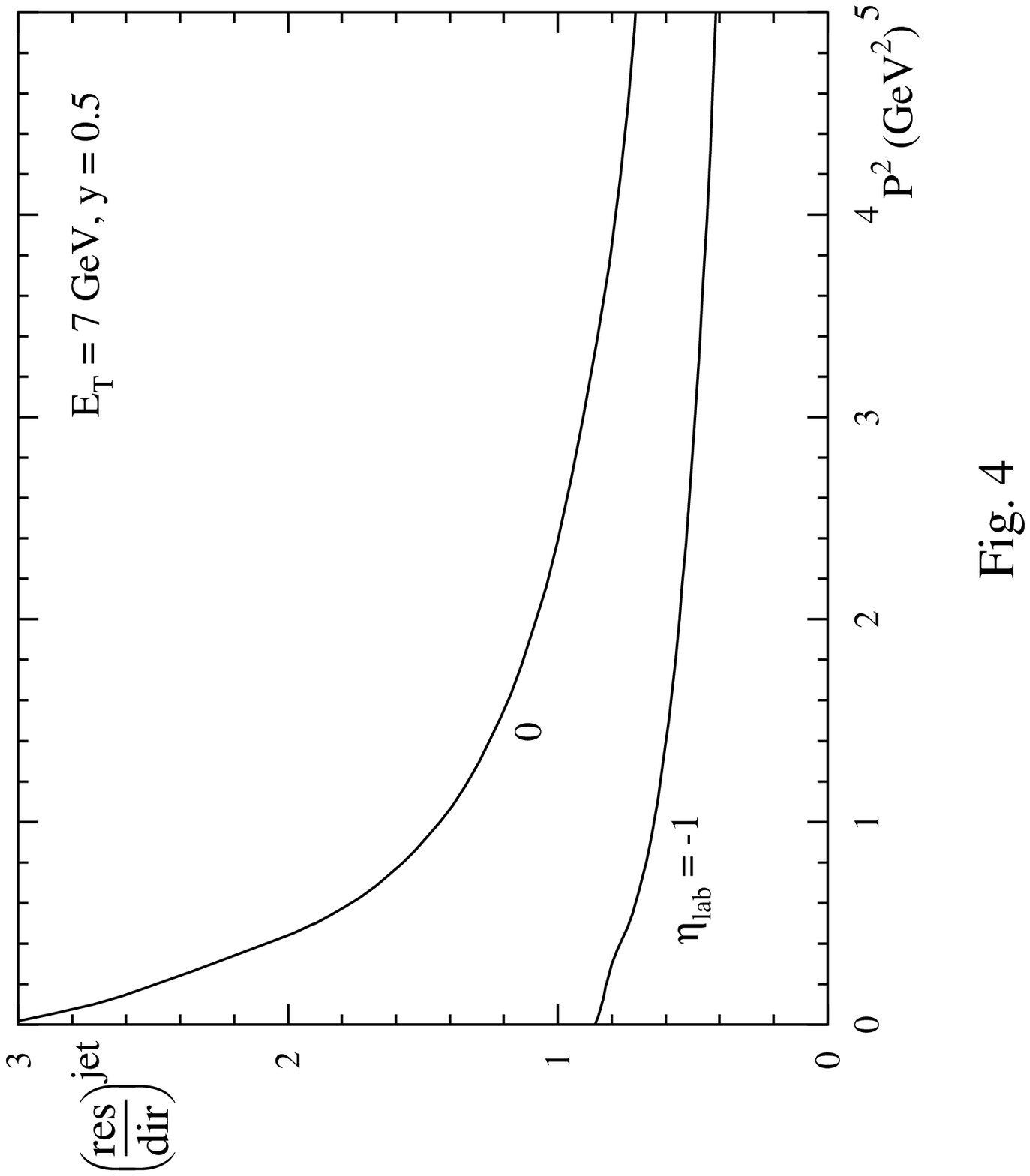,angle=90} 
\newpage 

\vspace*{-0.0cm}
\hspace*{-1.4cm}
\epsfig{file=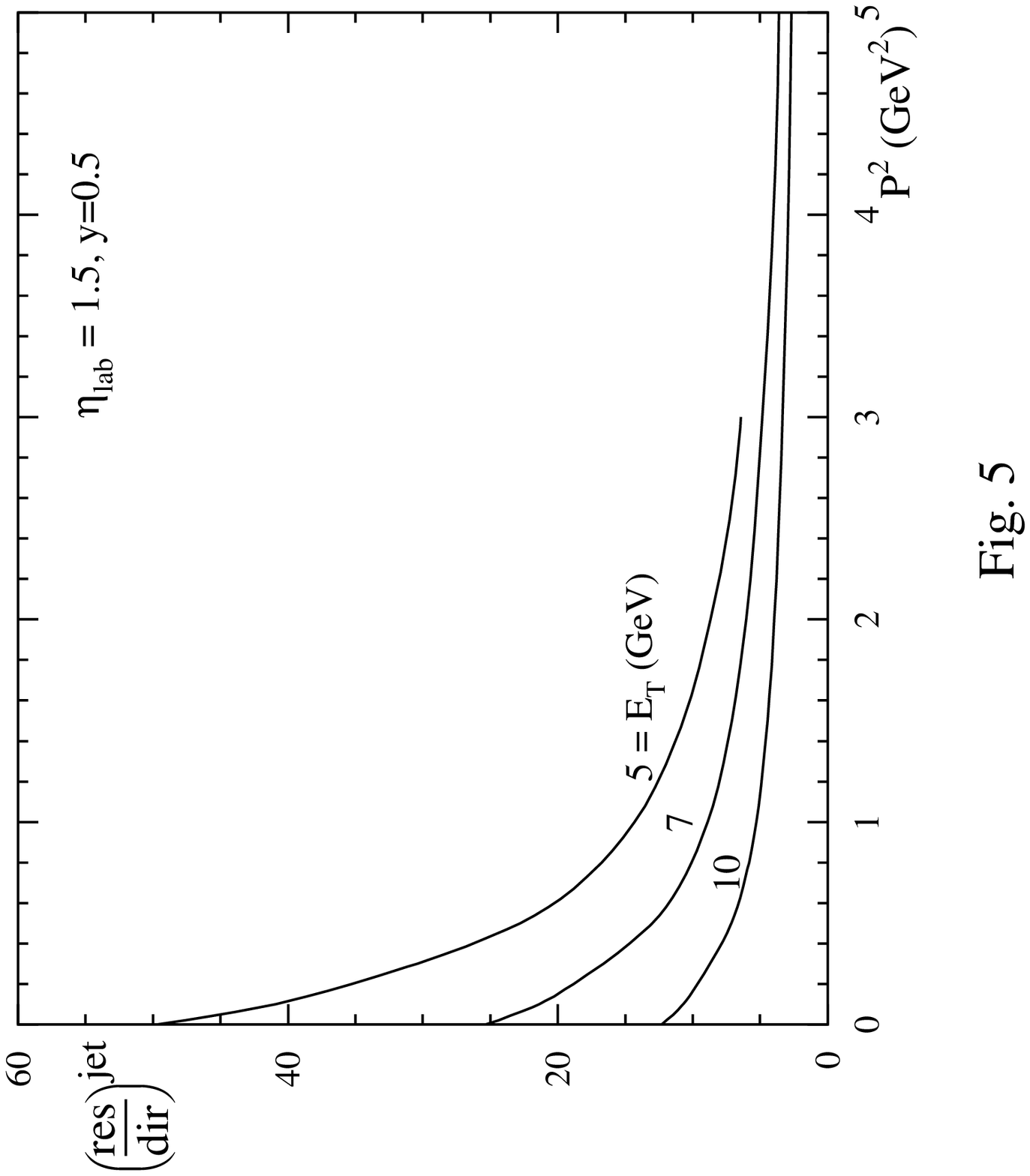,angle=90}
\newpage

\vspace*{-0.0cm}
\hspace*{-1.4cm}
\epsfig{file=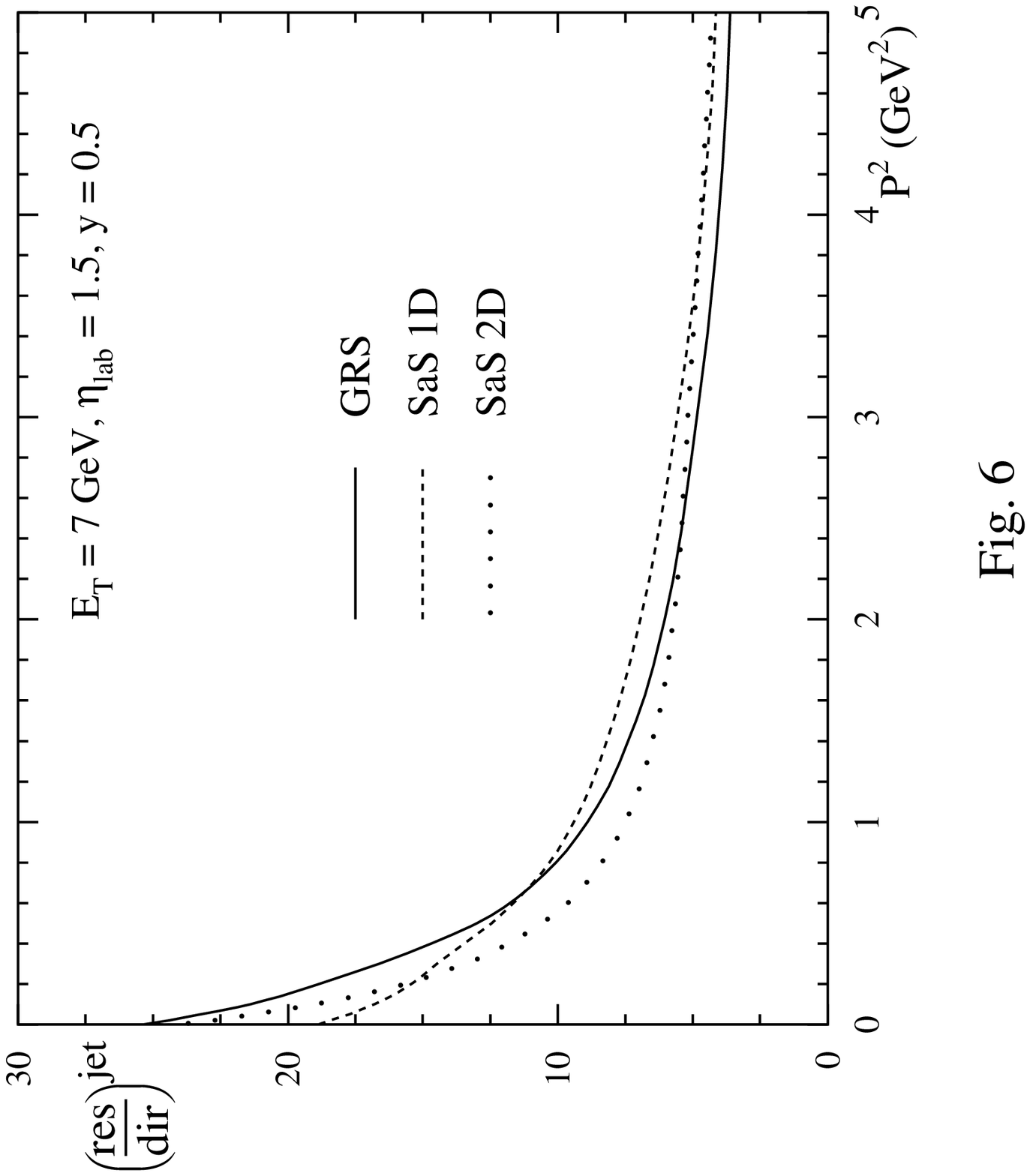,angle=90}

\begin{thebibliography}{99}
\bibitem{ref19} Higher values of $\eta_{lab}$ should be avoided in order
to guarantee clean jets, uncontaminated by the beam region.
\end{thebibliography}
\end{document}